\documentclass[12pt]{article}

\usepackage{amsmath}
\usepackage{amssymb}
\usepackage{graphicx}
\usepackage{amsthm}
\usepackage[round]{natbib}
\usepackage{multirow}
\usepackage{epstopdf}
\usepackage{xcolor}

\newtheorem{theorem}{Theorem}[section]
\newtheorem{corollary}{Corollary}[section]
\newtheorem{lemma}{Lemma}[section]


\numberwithin{equation}{section}

\title{Bayesian optimal designs for dose-response curves with common parameters}

\author{Kirsten Schorning, Maria Konstantinou\\
Fakult\"at f\"ur Mathematik \\
Ruhr-Universit\"at Bochum \\
44780 Bochum \\
Germany
}

\begin{document}

\date{}

\maketitle

\begin{abstract}

 The issue of determining not only an adequate dose but also a dosing frequency of a drug arises frequently in Phase II clinical trials. This results in the comparison of models which have some parameters in common. Planning such studies based on Bayesian optimal designs offers robustness to our conclusions since these designs, unlike locally optimal designs, are efficient even if the parameters are misspecified. In this paper we develop approximate design theory for Bayesian $D$-optimality for nonlinear regression models with common parameters and investigate the cases of common location or common location and scale parameters separately. Analytical characterisations of saturated Bayesian $D$-optimal designs are derived for frequently used dose-response models and the advantages of our results are illustrated via a numerical investigation.

\end{abstract}

%Keywords: Bayesian optimal designs, $D$-optimality, different treatment groups, nonlinear regression
%\smallskip

%AMS Subject Classification: 62K05 

\section{Introduction}

One of the key objectives in pharmaceutical drug development is to properly characterise the dose-response relationship as this has a direct impact on the determination of an efficacious and safe dose in Phase II clinical trials. As pointed out by \citet{dette08} and \citet{bretz10}, too high of a dose is likely to result in toxicity and safety problems whereas choosing too low of a dose reduces the chance of establishing adequate efficacy and thus the chance of approval of the drug. In fact, in a recent review \citet{sacks} identify the failure to select an optimal dose due to statistical uncertainty as one of the most frequent reasons for the delay or denial of approval of drugs by the Food and Drug Administration (FDA). 

The efficient planning of dose-finding studies can significantly improve the statistical accuracy for the selection of an efficacious and safe dose. For this reason numerous authors investigate the construction of optimal experimental designs for such studies and consider regression models which are frequently used to describe the dose-response relationship (see, for example, \citet{dette08}, \citet{dragalin} and \citet{bretz10}). 

In this paper we consider the closely related issue of determining not only a safe and efficacious dose but also the dosing frequency. This is often of interest in Phase II dose-finding studies since the frequency of administration affects the pharmacological effect of the drug under investigation. That is, if the same drug is administered at different dosing intervals, for example, once or twice a day, the drug exposure inside the body during the day will be higher right after administration and lower right before administration in the case of once a day administration whereas if the drug is administered twice a day the drug exposure during the day will be more uniform.

Under this scenario, estimating the dose-response curves corresponding to the different administration frequencies separately is inefficient. This is due to the fact that the regression models will often have common parameters thus suggesting a borrowing of strength. One of the parameters that it is reasonable to assume is the same for all models is the parameter corresponding to the placebo effect. Another assumption that is made in some situations is that there exists a biological maximum attainable drug effect. In this case, the parameter describing the efficacy for high doses can also be considered to be common for all models. In what follows we will investigate separately the cases where only the parameter corresponding to the placebo effect or both of the aforementioned parameters are common in all the regression models. 

The aim of the present paper is the efficient planning of such dose-finding studies where the regression models, used to model the dose-response relationship for the different administration frequencies, share some of the parameters. More specifically, we are interested in finding designs providing us with the doses to utilise in the different administration groups and also the way to split the total sample size between the groups, such that the design is optimal in terms of estimating all of the model parameters with high precision. Despite of the practical interest, the construction of optimal designs for dose-finding studies involving different administration frequencies has only been investigated by \citet{feller} who develop the optimal design theory corresponding to this scenario and derive explicit expressions for locally $D$-optimal designs. However, locally $D$-optimal designs require a best guess of the parameter values in order to be used in practice which can result in inefficient designs if the parameters are misspecified. Therefore, a design strategy that is robust to parameter misspecification is of great importance in applications.

Throughout this paper we adopt the robust design strategy of Bayesian optimal designs introduced by \citet{Pronzato85}, \citet{chaloner89} and \citet{ChalonerLarntz}. Given an uncertainty space for the parameter values, Bayesian optimality incorporates this uncertainty in the formulation of the optimality criteria via a prior distribution on the given parameter space and the resulting Bayesian optimality criteria are based on classical optimality criteria (see, for example, \citet{chaloner93} and \citet{Chaloner} for more details). The construction of Bayesian optimal designs for several regression models has been studied, for example, by \citet{dette2007}, \citet{dette08} and \citet{CHALONER1992}. However, to the best of the authors knowledge, Bayesian optimal designs for dose-finding studies involving several administration frequency groups have not been considered in the literature. 

In the present paper we consider the Bayesian $D$-optimal design problem for comparing regression models with common parameters. 
Since these designs are based on the results for the individual models, in Section \ref{secind} we derive analytical characterisations of Bayesian D-optimal designs for the individual regression models which, to the best of our knowledge, do not exist in the literature so far. In Section \ref{seccom} we introduce the regression models with common parameters and in Section \ref{seccomgen} we derive some general results concerning the Bayesian $D$-optimal designs. The scenarios where the regression models have only the location parameter or both the location and scale parameter in common are considered in Sections \ref{seccomloc} and \ref{seccomlocscale}, respectively. In both cases, we provide explicit characterisations of the saturated Bayesian $D$-optimal designs for regression models that are frequently used in dose-response studies. Finally, in Section \ref{secex} we asses the performance of the saturated designs constructed in Section \ref{seccom} in a situation where they are not globally optimal.

\section{Bayesian $D$-optimal designs for individual regression models} \label{secind}

It turns out that the locally optimal designs for models with common parameters are based on the locally optimal designs for the individual models. Therefore, we begin our investigation by considering the construction of Bayesian optimal designs for individual regression models. Besides some results for specific regression models, the construction of such designs in a general framework has not been considered in the literature so far.

\subsection{Models and optimality criterion}

We consider regression models of the form
\begin{equation}\label{ind-model}
Y_{jl} = f(d_j,\tilde{\boldsymbol{\theta}}) + \varepsilon_{jl}, \quad j =1, \ldots, \tilde{k}; l=1, \ldots, \tilde{n}_j, 
\end{equation}
where $\varepsilon_{jl}$ are independent centred normally distributed, that is, $\varepsilon_{jl} \overset{iid}{\sim} N(0, \sigma^2)$, \linebreak
$\tilde{\boldsymbol{\theta}} = (\theta_1, \ldots, \theta_{\tilde{m}})^T \in \tilde{\Theta} \subset \mathbb{R}^{\tilde{m}}$ is the vector of model parameters and $\tilde{n}_j$ is the total number of observations taken at each dose level $d_j$ ($j=1, \ldots, \tilde{k}$) taking values in the design space $\mathcal{X}=[0,d_{max}]$. We collect the information provided by the different dose levels in a probability measure, say $\tilde{\xi}$, on the design space $\mathcal{X}$ defined as
\begin{equation*}
\tilde{\xi} = 
\begin{Bmatrix}
d_1 & \ldots & d_{\tilde{k}} \\
\omega_1 & \ldots & \omega_{\tilde{k}}
\end{Bmatrix}, \qquad 
0 < \omega_j < 1, \qquad 
\sum _{j=1}^{\tilde{k}} \omega_j = 1 ,
\end{equation*}
(see \citet{kiefer}). The measure $\tilde\xi$ is called design, the $d_j$'s are called the support points of the design and the corresponding $\omega_j$'s are the weights of the design and represent the proportion of observations to be taken at each experimental condition $d_j$. We denote the support of the design $\tilde{\xi}$  by $\text{supp}(\tilde\xi)$.
Let $\tilde{n}=\sum_{j=1}^{\tilde{k}} \tilde{n}_j$ be the total sample size. Under the assumption
\begin{equation*}
\lim_{n \to \infty} \frac{\tilde{n}_j}{\tilde{n}} = \omega_j >0,  \qquad \qquad j=1, \dots, \tilde{k},
\end{equation*}
the maximum likelihood estimator $\hat{\tilde{\boldsymbol{\theta}}}$ of the parameter vector $\tilde{\boldsymbol{\theta}}$ satisfies
\begin{equation*}
\sqrt{\tilde{n}} (\hat{\tilde{\boldsymbol{\theta}}} - \tilde{\boldsymbol{\theta}}) \xrightarrow{\mathcal{D}} N_{\tilde{m}}(\boldsymbol{0},
M_{\text{ind}}^{-1}(\tilde{\xi}, \tilde{\boldsymbol{\theta}})  ), 
\end{equation*}
where 
\begin{equation}\label{info-ind}
M_{\text{ind}}^{-1}(\tilde{\xi}, \tilde{\boldsymbol{\theta}}) = \int_{\mathcal{X}} \tilde{h}(d,\tilde{\boldsymbol{\theta}}) \tilde{h}^T(d,\tilde{\boldsymbol{\theta}}) d\tilde{\xi}(d),
\end{equation}
and
\begin{equation}\label{deriv-ind}
\tilde{h}(d,\tilde{\boldsymbol{\theta}}) = \frac{1}{\sigma} \left(\frac{\partial}{\partial \tilde{\boldsymbol{\theta}}} f(d,\tilde{\boldsymbol{\theta}}) \right)^T.
\end{equation}
Our goal is to find optimal designs of this form such that the parameter vector $\tilde{\boldsymbol{\theta}}$ is estimated with high precision even when there is imperfect knowledge of the true parameter values.

As discussed in the introduction, throughout this paper we adopt the Bayesian approach as a robust design strategy and thus we address the issue of parameter dependence of locally optimal designs. Since interest is in estimating all of the model parameters, the Bayesian criterion is based on the classical $D$-optimality criterion (see \cite{Chaloner}
 for more details). In particular, if $\tilde{\pi}$ is a prior distribution on the parameter space $\tilde{\Theta}$, a design $\tilde{\xi}_{\tilde{\pi}}^*$ is called Bayesian $D$-optimal with respect to the prior $\tilde{\pi}$ for the individual model \eqref{ind-model} if it maximises the function
\begin{equation}\label{criterion-ind}
\Phi_{\tilde{\pi}}(\tilde{\xi}) = \int_{\tilde{\Theta}} \log \det\{M_{\text{ind}}(\tilde{\xi}, \tilde{\boldsymbol{\theta}})\} \,\tilde{\pi}(d \tilde{\boldsymbol{\theta}}).
\end{equation}
\subsection{Bayesian $D$-optimal designs}

A very important tool for characterising and checking the optimality of a candidate design is the general equivalence theorem. Using Theorem 3.3 in \citet{dette2007} the general equivalence theorem for Bayesian $D$-optimality of models of the form \eqref{ind-model} is given below. 
\begin{theorem}\label{GET-ind}
A design $\tilde{\xi}_{\tilde{\pi}}^*$ is Bayesian $D$-optimal with respect to the prior $\tilde{\pi}$ for maximum likelihood estimation in model \eqref{ind-model} if and only if the inequality
\begin{equation*}
\int_{\tilde{\Theta}} g(d,\tilde{\xi}_{\tilde{\pi}}^*,\tilde{\boldsymbol{\theta}}) \,\tilde{\pi}(d \tilde{\boldsymbol{\theta}}) := \int_{\tilde{\Theta}} h^T(d,\tilde{\boldsymbol{\theta}}) M_{\text{ind}}^{-1}(\tilde{\xi}_{\tilde{\pi}}^*,\tilde{\boldsymbol{\theta}}) h(d,\tilde{\boldsymbol{\theta}}) \tilde{\pi}(d \tilde{\boldsymbol{\theta}}) \leq \tilde{m},
\end{equation*}
holds for all $d \in \mathcal{X}=[0,d_{max}]$. Equality holds for any point $d \in \text{supp}(\tilde{\xi}_{\tilde{\pi}}^*)$.
\end{theorem}

An approximate design for maximum likelihood estimation in model \eqref{ind-model} must have at least $\tilde{m}$ support points in order  to be estimable for all the model parameters. A well known result in the optimal design theory is that of equally weighted $D$-optimal saturated designs, that is, of designs with as many support points as model parameters. Lemma \ref{equal-weights-ind} shows that this result holds also in the case of Bayesian $D$-optimal saturated designs for maximum likelihood estimation in model \eqref{ind-model} and it is proven in the appendix. 
\begin{lemma}\label{equal-weights-ind}
If $\tilde{\xi}_{\tilde{\pi}}^*$ is a Bayesian $D$-optimal saturated design with respect to the prior $\tilde{\pi}$ for maximum likelihood estimation in model \eqref{ind-model}, then it has equal weights $\frac{1}{\tilde{m}}$ at its support points.
\end{lemma}

For the rest of this section we will focus on three well-known nonlinear models which are widely used for the modelling of the dose-response relationship. Namely, we consider the Emax, the exponential and the linear-in-log models specified by the regression functions  
\begin{equation}\label{emax}
f_1(d,\tilde{\boldsymbol{\theta}}) = \theta_0 + \frac{\theta_1 d}{\theta_2 + d}, \quad d\in[0,d_{max}],
\end{equation}
\begin{equation}\label{exp}
f_2(d,\tilde{\boldsymbol{\theta}}) = \theta_0 + \theta_1 \left(\exp \left( \frac{d}{\theta_2}\right)-1 \right), \quad d\in[0,d_{max}],
\end{equation}
\begin{equation}\label{log}
f_3(d,\tilde{\boldsymbol{\theta}}) = \theta_0 + \theta_1 \log \left( \frac{d}{\theta_2}+1 \right), \quad d\in[0,d_{max}],
\end{equation}
respectively. The locally $D$-optimal designs for reparameterisations of these models are found in \citet{detkisbevbre2010} and are proven to be saturated, that is, they have exactly three support points, with the end-points of the design space always being two of the support points. Therefore, for each of the nonlinear models given above, we consider the construction of Bayesian $D$-optimal saturated designs for which we provide analytical characterisations presented in the following theorem. Its proof can be found in the appendix.
  
\begin{theorem}\label{designs-ind}
A Bayesian $D$-optimal saturated design with respect to the prior $\tilde{\pi}= \tilde{\pi}_0 \times \tilde{\pi}_1 \times \tilde{\pi}_2$ for maximum likelihood estimation in the Emax, exponential or the linear-in-log model is equally supported at points $0, \tilde{d}^*, d_{max}$ where the non-trivial support point $\tilde{d}^*$ is a solution in the interval $(0,d_{max})$ of the equation
\begin{equation}\label{eq-emax-ind}
\int_{\tilde{\Theta}_2}  \left( \frac{1}{d} - \frac{1}{d_{max}-d} - \frac{2}{\theta_2 + d} \right) \,\tilde{\pi}_{2}(d \theta_2) = 0, 
\end{equation}
for the Emax model, of the equation
\begin{equation} \label{eq-exp-ind}
\int_{\tilde{\Theta}_2}  \frac{\exp(d/\theta_2) \left(d_{\max} \exp(d_{\max}/\theta_2) - ( d + \theta_2)(\exp(d_{\max}/\theta_2) - 1) \right) }{d_{\max} \exp(d_{\max}/\theta_2) (\exp(d/\theta_2) -1 )- d \exp(d/\theta_2) (\exp(d_{\max}/\theta_2) -1 )}\,\tilde{\pi}_{2}(d \theta_2) =0
\end{equation}
for the exponential model and of the equation
\begin{equation} \label{eq-log-ind}
\int_{\tilde{\Theta}_2} \frac{1}{d+\theta_2} \frac{\theta_2 \log(d_{\max}/\theta_2 +1) (d_{\max} + \theta_2) - d_{\max} (d+ \theta_2)}{d(d_{\max}+ \theta_2) \log(d_{\max}/\theta_2 +1) - d_{\max}(d+\theta_2) \log(d/\theta_2 +1)} \,\tilde{\pi}_{2}(d \theta_2) = 0, 
\end{equation}
for the linear-in-log model.
\end{theorem}

Note that these analytical characterisations reduce the design problem to solving an equation in one variable and therefore, the numerical effort for design search is reduced substantially. Furthermore, the Bayesian $D$-optimal designs require only the marginal prior distribution $\tilde{\pi}_{2}$ on $\tilde{\Theta}_2$, that is, they require prior information only for the parameter $\theta_2$ and not for the full parameter vector, because the considered models are linear in the first two parameters $\theta_0$ and $\theta_1$. 

%{\bf Are we sure we can not find un upper bound for the number of support points? Also can we motivate the use of saturated designs a bit better? Perhaps in the application section use more than three support points and see how much, or rather how little, we gain by adding support points. The same for designs of section 3?}

\section{Bayesian $D$-optimal designs for regression models with common parameters}\label{seccom}

In this section we focus on our core investigation regarding the case of $M$ different administration frequency groups  modelled using the same parametric form %{\bf explanation?} 
and as it is discussed in the introduction, these models could share some parameters while others are different. More precisely, we consider regression models of the form
\begin{equation}\label{models}
Y_{ijl} = f(d_j^{(i)},\boldsymbol{\theta_1},\boldsymbol{\theta_2^{(i)}}) + \varepsilon_{ijl}, \quad i =1, \ldots, M; j=1,\ldots, k_i; l=1, \ldots, n_{ij} ,
\end{equation}
where $\varepsilon_{ijl} \overset{iid}{\sim} N(0, \sigma^2_i)$ and in each administration frequency group $i=1, \ldots, M$ a total of $n_{ij}$ observations are taken at each dose level $d_1^{(i)}, \ldots, d_{k_i}^{(i)}$. We further assume that the dose levels $d_j^{(i)}$, $i =1, \ldots, M; j=1,\ldots, k_i$, vary in possibly different design spaces for each group, that is, $\mathcal{X}_i = [0, d_{max}^{(i)}]$, $i=1, \ldots, M$. The vector $\boldsymbol{\theta_1} = (\theta_1, \ldots, \theta_p)^T \in \Theta_1 \subset \mathbb{R}^p$ corresponds to the parameters that are the same in all groups whereas $\boldsymbol{\theta_2^{(i)}} = (\theta_1^{(i)}, \ldots, \theta_q^{(i)})^T \in \Theta_2^{(i)} \subset \mathbb{R}^q$, $i=1, \ldots, M$, is the vector of parameters that are different for different groups.

\subsection{General results}\label{seccomgen}

Planning a dose-finding study with different administration frequency groups is concerned with identifying the doses to be utilised in each group and also identifying how to split the total sample size between the groups. We formulate this using approximate designs in the sense of \citet{kiefer}. In particular, for each group $i=1, \ldots, M$, we define the approximate design
\begin{equation*}
\xi_{i} = \begin{Bmatrix} d_1^{(i)} & \ldots & d_{k_i}^{(i)} \\ \omega_{i1} & \ldots & \omega_{ik_i} \end{Bmatrix}
,\qquad
0 < \omega_{ij} < 1
,\qquad
\sum_{j=1}^{k_i} \omega_{ij}=1, 
\end{equation*}
as a probability measure on the design space $\mathcal{X}_i$ assigning a proportion $\omega_{ij}$ of the observations in the $i$th group at each of the dose levels $d_j^{(i)}$ ($i =1, \ldots, M; j=1,\ldots, k_i$). Furthermore, we define the probability measure 
\begin{equation*}
\mu = \begin{Bmatrix} 1 & \ldots & M \\ \lambda_{1} & \ldots & \lambda_M \end{Bmatrix}
,\qquad
0 < \lambda_i < 1
,\qquad
\sum_{i=1}^{M} \lambda_i=1,
\end{equation*}
on the set $\{1,\ldots, M \}$ assigning a proportion $\lambda_i$ of the total sample size to the $i$th group. The collection of these designs in the vector $\xi=(\xi_1, \ldots, \xi_M, \mu)$ is a probability measure on the design space $\mathcal{X}_1 \times \ldots \mathcal{X}_M \times \{1,\ldots, M \}$ and is also called a design throughout this paper.

As discussed in the introduction, estimating the dose-response curves corresponding to each group separately, using the results of section 2, is wasteful because of the borrowing of strength due to the common parameters in the models. We therefore consider the simultaneous maximum likelihood estimation of the full parameter vector $\boldsymbol{\theta}=(\boldsymbol{\theta_1}, \boldsymbol{\theta_2^{(1)}}, \ldots, \boldsymbol{\theta_2^{(M)}})^T \in \Theta_1 \times \Theta_2^{(1)} \times \ldots \Theta_2^{(M)}=\Theta \subset \mathbb{R}^m$ where $m=p + q M$. 

Let $n_i = \sum_{j=1}^{k_i} n_{ij}$ be the sample size in the $i$th group and denote by $n=\sum_{i=1}^M n_{i}$ the total sample size. From \citet{feller} we have that under common regularity assumptions and the assumptions 
\begin{equation*}
\lim_{n_i \to \infty} \frac{n_{ij}}{n_i} = \omega_{ij} \qquad \text{and} \qquad  \lim_{n \to \infty} \frac{n_{i}}{n} = \lambda_{i} \qquad (i=1, \ldots, M ; j=1, \ldots, k_i) , 
\end{equation*}
the maximum likelihood estimator $\hat{\boldsymbol{\theta}}$ of the full parameter vector $\boldsymbol{\theta}$ satisfies
\begin{equation*}
\sqrt{n} (\hat{\boldsymbol{\theta}} - \boldsymbol{\theta}) \xrightarrow{\mathcal{D}} N_{m}(\boldsymbol{0},
M^{-1}(\xi, \boldsymbol{\theta})  ),
\end{equation*}
where 
\begin{equation}\label{info}
M(\xi, \boldsymbol{\theta}) = \int \int_{\mathcal{X}_z} h_z(d,\boldsymbol{\theta}) h_z^T(d,\boldsymbol{\theta}) d\xi_z(d) d\mu(z) = \sum_{i=1} ^ M \lambda_i M^{(i)}(\xi_i, \boldsymbol{\theta}),
\end{equation}
is the information matrix of the design $\xi = (\xi_1, \ldots, \xi_M, \mu)$, the matrices $M^{(i)}$ are defined as
\begin{equation}\label{info-small}
M^{(i)}(\xi_i, \boldsymbol{\theta}) = \int_{\mathcal{X}_i} h_i(d,\boldsymbol{\theta}) h_i^T(d,\boldsymbol{\theta}) d\xi_i(d), \qquad i=1, \ldots, M ,
\end{equation}
and
\begin{equation}\label{deriv}
h_i(d,\boldsymbol{\theta}) = \frac{1}{\sigma_i} \left( \left(\frac{\partial}{\partial \boldsymbol{\theta}_1} f(d,\boldsymbol{\theta}_1, \boldsymbol{\theta}_2^{(i)}) \right)^T, \boldsymbol{0}^T_{q(i-1)}, \left(\frac{\partial}{\partial \boldsymbol{\theta}_2^{(i)}} f(d,\boldsymbol{\theta}_1, \boldsymbol{\theta}_2^{(i)}) \right)^T, \boldsymbol{0}^T_{q(M-i)} \right)^T \in \mathbb{R}^m ,
\end{equation}
is the gradient of the regression function $f(d,\boldsymbol{\theta}_1, \boldsymbol{\theta}_2^{(i)})$ with respect to the parameter vector $\boldsymbol{\theta} \in \mathbb{R}^m$, with $\boldsymbol{0}_{q(i-1)} \in \mathbb{R}^{q(i-1)}$ and $\boldsymbol{0}_{q(M-i)} \in \mathbb{R}^{q(M-i)}$ {denoting} vectors with all entries equal to zero.

Let $\pi$ be a prior distribution on the parameter space $\Theta = \Theta_1 \times \Theta_2^{(1)} \times \ldots \Theta_2^{(M)}$. A design $\xi_{\pi}^*$ is Bayesian $D$-optimal with respect to the prior $\pi$ for maximum likelihood estimation in models \eqref{models} if it maximises
\begin{equation}\label{criterion}
\Phi_{\pi}(\xi) = \int_{\Theta} \log \det \{ M(\xi, \boldsymbol{\theta}) \} \pi(d \boldsymbol{\theta}).
\end{equation} 
The following theorem provides the general equivalence theorem for characterising and checking the Bayesian $D$-optimality of a candidate design in the multiple administration frequency groups scenario we consider. Its proof is given in the appendix.

\begin{theorem}\label{get}
A design $\xi^*_{\pi} = (\xi_1^*, \ldots, \xi_M^*,\mu^*)$ is Bayesian $D$-optimal with respect to the prior $\pi$ for maximum likelihood estimation in models \eqref{models}, if and only if the $M$ inequalities
\begin{equation}\label{tau_get}
\tau_{i}(d) = \int_{\Theta} g_i(d,\xi^*_{\pi},\boldsymbol{\theta}) \pi(d \boldsymbol{\theta}) - m:= \int_{\Theta} h_i^T(d,\boldsymbol{\theta}) M^{-1}(\xi^*_{\pi}, \boldsymbol{\theta}) h_i(d,\boldsymbol{\theta}) \pi(d \boldsymbol{\theta}) - (p+qM) \leq 0,
\end{equation}
are satisfied for all $d \in \mathcal{X}_i$, $i=1, \ldots, M$. Equality holds for any points $(d_1, \ldots, d_M, z) \in \text{supp}(\xi_1^*) \times \ldots \times \text{supp}(\xi_M^*) \times \text{supp}(\mu^*)$.
\end{theorem}

Now let
\begin{equation}\label{class}
\Xi_m^M = \left\{ \xi_{\pi} = (\xi_1, \ldots, \xi_M, \mu) | \sum_{i=1}^M |\text{supp}(\xi_i)| = m \right\},
\end{equation}
be the set of all designs on $\mathcal{X}_1 \times \ldots \times \mathcal{X}_M \times  \{1,\ldots,M \}$,  with a total of exactly $m$ different dose levels in the $M$ groups. Therefore, this is the class of saturated designs on $\mathcal{X}_1 \times \ldots \times \mathcal{X}_M \times  \{1,\ldots,M \}$. Lemma \ref{equal-weights} provides us with the weights of Bayesian $D$-optimal designs in the class $\Xi_m^M$ and is proven in the appendix.

\begin{lemma}\label{equal-weights}
Let $\xi_{\pi} = (\xi_1, \ldots, \xi_M, \mu) \in \Xi_m^M$ denote a design on $\mathcal{X}_1 \times \ldots \times \mathcal{X}_M \times  \{1,\ldots,M \}$ and $m_i$ denote the number of support points of $\xi_i$ ($i=1,\ldots,M$). Also assume that the $m=\sum_{i=1}^M m_i$ vectors $h_1(d_1^{(1)},\boldsymbol{\theta}), \ldots, h_1(d_{m_1}^{(1)},\boldsymbol{\theta}), \ldots, h_M(d_{1}^{(M)},\boldsymbol{\theta}), \ldots, h_M(d_{m_M}^{(M)},\boldsymbol{\theta})$ are linearly independent for all parameters $\boldsymbol{\theta} \in \Theta$, where $d_j^{(i)} \in \text{supp}(\xi_i)$, $j=1, \ldots, m_i; i=1, \ldots, M$.

If $\xi_{\pi}^*=(\xi_1^*, \ldots, \xi_M^*, \mu^*)$ is Bayesian $D$-optimal with respect to the prior $\pi$ in the class $\Xi_m^M$, then each $\xi_i^*$ has equal weights at its support points and the weights of $\mu^*$ at its support points $1, \ldots, M$ are given by $\frac{m_1}{m}, \ldots, \frac{m_M}{m}$ respectively.
\end{lemma}

In what follows we focus on the construction of designs which are Bayesian $D$-optimal in the class $\Xi_m^M$, that is, of designs which are saturated. This is because these designs have the smallest possible number of support points and thus, the number of experimental conditions where observations should be taken is minimum. Therefore, saturated designs result in lower costs for the study and are thus often preferred to designs which are optimal amongst all possible designs if the latter have more support points (see Section 4 for more details).

Moreover, in the following two sections we investigate the cases of the models \eqref{models} having the same location parameter and having the same location and scale parameters separately. Table \ref{table} presents these two cases with respect to the forms of the regression functions for the Emax, exponential and linear-in-log models.

\begin{table}[h!]\label{table}
\caption{\it{Emax, exponential and linear-in-log models for $i=1, \ldots, M$ groups. Left column: Common location parameter. Right column: Common location and scale parameters.}}

\vspace{0.3cm}

\centering
\begin{tabular}{|ccc|}

\hline

 & location  &  location and scale  \\

%\hline

Emax  & $\theta_1 + \theta_1^{(i)}\frac{d}{\theta_2^{(i)}+d}$ & $\theta_1 + \theta_2\frac{d}{\theta_2^{(i)}+d}$    \\
Exponential  & $\theta_1 + \theta_1^{(i)} \Big( \exp \big( \frac{d}{\theta_2^{(i)}} \big) -1 \Big)$ & $\theta_1 + \theta_2 \Big( \exp \big( \frac{d}{\theta_2^{(i)}} \big) -1 \Big)$    \\
Linear-in-log  & $\theta_1 + \theta_1^{(i)} \log \Big( \frac{d}{\theta_2^{(i)}} + 1 \Big)$ & $\theta_1 + \theta_2 \log \Big( \frac{d}{\theta_2^{(i)}} + 1 \Big)$   \\

\hline

\end{tabular}
\end{table}

\subsection{Bayesian $D$-optimal designs for models with the same location parameter}\label{seccomloc}

We first consider the case where the regression models corresponding to the $M$ administration frequency groups share only the location parameter and thus, the regression function in \eqref{models} is of the form
\begin{equation}\label{models-location}
f(d, \theta_1, \boldsymbol{\theta}_2^{(i)}) = \theta_1 + f_0(d, \boldsymbol{\theta}_2^{(i)}), \qquad i=1, \ldots, M.
\end{equation}
This corresponds, for example, to the case of a common placebo effect for all groups. Here the full parameter vector is $\boldsymbol{\theta}=(\theta_1, \boldsymbol{\theta}_2^{(1)}, \ldots, \boldsymbol{\theta}_2^{(M)})^T \in \Theta \subset \mathbb{R}^m$ where $m=1+qM$.

Denote by $\tilde{\pi}^{(i)}_2$ the marginal prior distribution on $\Theta_2^{(i)}$ ($i=1, \ldots, M$). The following theorem provides a solution of the Bayesian $D$-optimal design problem in the class $\Xi_m^M$ for models \eqref{models} with regression functions \eqref{models-location} if the Bayesian $D$-optimal saturated designs for the individual models are known. Its proof can be found in the appendix.

\begin{theorem}\label{designs-location-general}
Let $\sigma_1^2 = \min_{i=1, \ldots, M} \sigma_i^2$ and consider models \eqref{models} with regression functions of the form \eqref{models-location} also satisfying
\begin{equation}\label{cond32}
f_0(0, \boldsymbol{\theta}_2^{(i)}) = 0, \qquad \eta_0(0, \boldsymbol{\theta}_2^{(i)}) = \frac{\partial}{\partial \boldsymbol{\theta}_2^{(i)}} f_0(d, \boldsymbol{\theta}_2^{(i)}) |_{d=0} = \boldsymbol{0}_q, \qquad (i=1, \ldots, M).
\end{equation}
If the saturated design 
\begin{equation}\label{individualdesign}
 \tilde{\xi}_{\tilde{\pi}_2^{(i)}}^{*} = \begin{Bmatrix} 0 & d_1^{(i)} & \ldots &  d_q^{(i)}  \\ \frac{1}{q+1} & \frac{1}{q+1} & \ldots & \frac{1}{q+1} \end{Bmatrix},
\end{equation}
is the Bayesian $D$-optimal saturated design with respect to the prior $\tilde{\pi}^{(i)}_2$ on $\Theta_2^{(i)}$ for maximum likelihood estimation in the individual model with variance $\sigma_i^2$ ($i=1, \ldots, M$), then the Bayesian $D$-optimal design in the class $\Xi_5^2$ with respect to the prior $\pi = \tilde{\pi}_1 \times \tilde{\pi}^{(1)}_2 \times \ldots \times \tilde{\pi}^{(M)}_2$ for maximum likelihood estimation in models \eqref{models} with regression function \eqref{models-location} is given by $\xi_{\pi}^* = (\xi_1^*, \ldots, \xi_M^*, \mu^*)$, where
\begin{equation}\label{32alldes}
\xi_{1}^* = \tilde{\xi}_{\tilde{\pi}_2^{(1)}}^{*}; \qquad \xi_i^* = \begin{Bmatrix} d_1^{(i)} & \ldots & d_q^{(i)} \\ \frac{1}{q} & \ldots &  \frac{1}{q} \end{Bmatrix}, \quad i=2, \ldots, M; \qquad \mu^* = \begin{Bmatrix} 1 & 2 & \ldots & M \\ \frac{q+1}{m} & \frac{q}{m} & \ldots &  \frac{q}{m} \end{Bmatrix}.
\end{equation}
\end{theorem}

Therefore, the common location parameter ${\bf \theta_1}$ is estimated in the group with minimal variance. Intuitively this result is reasonable since the estimator of the common location parameter ${\bf \theta_1}$ is roughly the mean of the observations at dose level $d=0$ and the variance of this estimator is proportional to the variance in the considered group.

The corresponding Bayesian $D$-optimal saturated designs for the cases of Emax, exponential and linear-in-log models are found explicitly and are presented in Corollary \ref{designs-location}. This result follows by a direct application of Theorem \ref{designs-location-general} using the corresponding characterisations presented in Theorem \ref{designs-ind} for Bayesian $D$-optimal saturated designs in the individual Emax, exponential and linear-in-log models and thus, its proof is omitted. 

\begin{corollary}\label{designs-location}
Let $\sigma_1^2 = \min_{i=1, \ldots, M} \sigma_i^2$. The Bayesian $D$-optimal design in the class \eqref{class} with respect to the prior $\pi = \tilde{\pi}_1 \times \tilde{\pi}^{(1)}_2 \times \ldots \times \tilde{\pi}^{(M)}_2$  for maximum likelihood estimation in Emax, exponential and linear-in-log models \eqref{models} with regression functions \eqref{models-location} is $\xi_{\pi}^* = (\xi_1^*, \ldots, \xi_M^*, \mu^*)$ where 
\begin{equation*}
\xi_1^* = \begin{Bmatrix} 0 & d^{*,(1)} & d_{max}^{(1)} \\ \frac{1}{3} & \frac{1}{3} & \frac{1}{3} \end{Bmatrix} 
;\quad \xi_i^* = \begin{Bmatrix}  d^{*,(i)} & d_{max}^{(i)} \\ \frac{1}{2} & \frac{1}{2}  \end{Bmatrix}, \quad i=2, \ldots, M;\quad \mu^* = \begin{Bmatrix}  1 & 2 & \ldots & M \\ \frac{3}{m} & \frac{2}{m} & \ldots & \frac{2}{m} \end{Bmatrix},
\end{equation*}
and the point $\tilde{d}^{*,(i)}$ ($i=1, \ldots, M$) is a solution of the equation \eqref{eq-emax-ind}, \eqref{eq-exp-ind} and \eqref{eq-log-ind} for the cases of Emax, exponential and linear-in-log individual models respectively using the corresponding marginal prior $\tilde{\pi}^{(i)}_2$. 
\end{corollary}

\subsection{Bayesian $D$-optimal designs for models with the same location and scale parameters}\label{seccomlocscale}

In this section we consider models \eqref{models} with regression functions of the form 
\begin{equation}\label{models-locationscale}
f(d, \theta_1, \theta_2, \boldsymbol{\theta}_2^{(i)}) = \theta_1 + \theta_2 f_0(d, \boldsymbol{\theta}_2^{(i)}), \qquad i=1, \ldots, M,
\end{equation}
that is, the models have common location and scale parameters. One such situation is, for example, when both the placebo effect and the efficacy for high doses coincide across the different models. As noted by \citet{feller}, the optimal design problem in this case is significantly harder. Therefore, in what follows we will present results for the case $M=2$ administration frequency groups and thus the full parameter vector is $\boldsymbol{\theta}=(\theta_1, \theta_2, \theta_2^{(1)}, \theta_2^{(2)})^T \in \Theta \subset \mathbb{R}^4$. Similar results can be obtained for $M>2$ with an additional amount of notation. 

For the rest of this section we will focus on Emax, exponential and linear-in-log models for which we derive analytical characterisations of their corresponding Bayesian $D$-optimal designs. We also note that we construct Bayesian $D$-optimal designs in the class $\Xi_4^2$, defined in \eqref{class}, of saturated designs on $\mathcal{X}_1 \times \mathcal{X}_2 \times \{1, 2\}$. We begin with the explicit result for two Emax models with the same location and scale parameters. The proof can be found in the appendix.

\begin{theorem}\label{designs-emax}
Let $r=\frac{\sigma_1^2}{\sigma_2^2}$ and $\tilde{d}^{*,(i)}$, $d^{*,(i)}$ be solutions of the equation \eqref{eq-emax-ind} in the interval $(0,d_{max}^{(i)})$ and of the equation
\begin{equation*}
\int_{\Theta_2^{(i)}} \frac{1}{d} - \frac{2}{d+\theta_2^{(i)}}  \tilde{\pi}_2^{(i)}(d \theta_2^{(i)}) = 0,
\end{equation*}
in the interval $[0,d_{max}^{(i)}]$ respectively using in both equations the marginal prior distribution $\tilde{\pi}_2^{(i)}$ on $\Theta_2^{(i)}$ for $i=1,2$. Also define the function
\begin{align*}
&u(\tilde{d}^{*,(1)}, d^{*,(1)}, \tilde{d}^{*,(2)}, d^{*,(2)}) := 2 \log \left[ \frac{\tilde{d}^{*,(1)} d_{max}^{(1)} (d_{max}^{(1)} - \tilde{d}^{*,(1)}) d^{*,(2)}}{\tilde{d}^{*,(2)} d_{max}^{(2)} (d_{max}^{(2)} - \tilde{d}^{*,(2)}) d^{*,(1)}} \right] \\
&+ 4 \sum_{i=1}^{2} (-1)^{i+1} \int_{\Theta_2^{(i)}}  \log \left[ \frac{(d^{*,(i)} + \theta_2^{(i)})}{(\tilde{d}^{*,(i)} + \theta_2^{(i)})(d_{max}^{(i)}+\theta_2^{(i)})} \right] \tilde{\pi}_2^{(i)}(d \theta_2^{(i)}).
\end{align*}
The Bayesian $D$-optimal design in the class $\Xi_4^2$ with respect to the prior $\pi= \tilde{\pi}_1 \times \tilde{\pi}_2 \times \tilde{\pi}_2^{(1)} \times \tilde{\pi}_2^{(2)}$ for maximum likelihood estimation in two Emax models with regression functions of the form \eqref{models-locationscale} is 
\begin{enumerate}
\item[(1)] $\xi^{a,*}=(\xi_1^{a,*}, \xi_2^{a,*}, \mu^{a,*})$ if $\log r \leq 0$ and $\log r \leq u(\tilde{d}^{*,(1)}, d^{*,(1)}, \tilde{d}^{*,(2)}, d^{*,(2)}) $, where
\begin{equation*}
\xi_1^{a,*} = \begin{Bmatrix} 0 & \tilde{d}^{*,(1)} & d_{max}^{(1)} \\ \frac{1}{3} & \frac{1}{3} & \frac{1}{3} \end{Bmatrix}, \quad \xi_2^{a,*} = \begin{Bmatrix} d^{*,(2)} \\ 1 \end{Bmatrix}, \quad \mu^{a,*} = \begin{Bmatrix} 1 & 2 \\ \frac{3}{4} & \frac{1}{4} \end{Bmatrix},
\end{equation*}
\item[(2)] $\xi^{b_1,*}=(\xi_1^{b_1,*}, \xi_2^{b_1,*}, \mu^{b_1,*})$ if $u(\tilde{d}^{*,(1)}, d^{*,(1)}, \tilde{d}^{*,(2)}, d^{*,(2)}) < \log r \leq 0$, where
\begin{equation*}
\xi_1^{b_1,*} = \begin{Bmatrix} 0 & d^{*,(1)} \\ \frac{1}{2} & \frac{1}{2} \end{Bmatrix}, \quad \xi_2^{b_1,*} = \begin{Bmatrix} \tilde{d}^{*,(2)} & d_{max}^{(2)} \\ \frac{1}{2} & \frac{1}{2} \end{Bmatrix}, \quad \mu^{b_1,*} = \begin{Bmatrix} 1 & 2 \\ \frac{1}{2} & \frac{1}{2} \end{Bmatrix},
\end{equation*}
\item[(3)] $\xi^{b_2,*}=(\xi_1^{b_2,*}, \xi_2^{b_2,*}, \mu^{b_2,*})$ if $0 \leq \log r \leq u(\tilde{d}^{*,(1)}, d^{*,(1)}, \tilde{d}^{*,(2)}, d^{*,(2)})$, where
\begin{equation*}
\xi_1^{b_2,*} = \begin{Bmatrix} \tilde{d}^{*,(1)} & d_{max}^{(1)} \\ \frac{1}{2} & \frac{1}{2} \end{Bmatrix}, \quad \xi_2^{b_2,*} = \begin{Bmatrix} 0 & d^{*,(2)} \\ \frac{1}{2} & \frac{1}{2} \end{Bmatrix}, \quad \mu^{b_2,*} = \begin{Bmatrix} 1 & 2 \\ \frac{1}{2} & \frac{1}{2} \end{Bmatrix},
\end{equation*}
\item[(4)] $\xi^{c,*}=(\xi_1^{c,*}, \xi_2^{c,*}, \mu^{c,*})$ if $\log r\geq0 $ and $\log(r) > u(\tilde{d}^{*,(1)}, d^{*,(1)}, \tilde{d}^{*,(2)}, d^{*,(2)})$, where
\begin{equation*}
\xi_1^{c,*} = \begin{Bmatrix} d^{*,(1)} \\ 1 \end{Bmatrix}, \quad \xi_2^{c,*} = \begin{Bmatrix} 0 & \tilde{d}^{*,(2)} & d_{max}^{(2)} \\ \frac{1}{3} & \frac{1}{3} & \frac{1}{3} \end{Bmatrix}, \quad \mu^{c,*} = \begin{Bmatrix} 1 & 2 \\ \frac{1}{4} & \frac{3}{4} \end{Bmatrix}.
\end{equation*}
\end{enumerate}
\end{theorem}

The analogous results for the cases of two exponential and two linear-in-log models with the same location and scale parameters are presented in the following theorems. Their proofs follow along the same lines as the proof of Theorem \ref{designs-emax} and are therefore omitted. 

\begin{theorem}\label{designs-exp}
Let $r=\frac{\sigma_1^2}{\sigma_2^2}$ and $\tilde{d}^{*,(i)}$ be a solution in the interval $(0,d_{max}^{(i)})$ of the equation \eqref{eq-exp-ind} using the marginal prior distribution $\tilde{\pi}_2^{(i)}$ on $\Theta_2^{(i)}$. Also define the function
%\begin{align*}
%&u(\tilde{d}^{*,(1)}, \tilde{d}^{*,(2)}) := \int_{\Theta_2^{(1)}} 2 \log \Bigg[ \exp\Big(\frac{\tilde{d}^{*,(1)}+ d_{max}^{(1)}}{\theta_2^{(1)}}\Big) (\tilde{d}^{*,(1)} - d_{max}^{(1)}) + d_{max}^{(1)} \exp \Big( \frac{d_{max}^{(1)}}{\theta_2^{(1)}} \Big) - \tilde{d}^{*,(1)} \exp \Big( \frac{\tilde{d}^{*,(1)}}{\theta_2^{(1)}} \Big) \Bigg] \\
%&- \frac{2}{\theta_2^{(1)}} \tilde{\pi}_2^{(1)}(d \theta_2^{(1)}) + \int_{\Theta_2^{(2)}} \frac{2}{\theta_2^{(2)}} - 2 \log \Bigg[ \exp\Big(\frac{\tilde{d}^{*,(2)}+ d_{max}^{(2)}}{\theta_2^{(2)}}\Big) (\tilde{d}^{*,(2)}- d_{max}^{(2)}) + d_{max}^{(2)} \exp \Big( \frac{d_{max}^{(2)}}{\theta_2^{(2)}} \Big) \\
%& - \tilde{d}^{*,(2)} \exp \Big( \frac{\tilde{d}^{*,(2)}}{\theta_2^{(2)}} \Big) \Bigg] \tilde{\pi}_2^{(2)}(d \theta_2^{(2)}).
%\end{align*}
\begin{align*}
&u(\tilde{d}^{*,(1)}, \tilde{d}^{*,(2)}) \\
& := 2 \sum_{i=1}^{2} (-1)^{i+1} \int_{\Theta_2^{(i)}} 2 \log \Bigg[ \Big(1- \tfrac{\tilde{d}^{*,(i)}}{ d_{max}^{(i)} } \Big)\exp\Big(\tfrac{\tilde{d}^{*,(i)}}{\theta_2^{(i)}}\Big) +  \tfrac{\tilde{d}^{*,(i)}}{ d_{max}^{(i)}} \exp \Big( \tfrac{\tilde{d}^{*,(i)}- d_{max}^{(i)}}{\theta_2^{(1)}} \Big) -1\Bigg]  \tilde{\pi}_2^{(i)}(d \theta_2^{(i)}) 
\end{align*}

The Bayesian $D$-optimal design in the class $\Xi_4^2$ with respect to the prior $\pi= \tilde{\pi}_1 \times \tilde{\pi}_2 \times \tilde{\pi}_2^{(1)} \times \tilde{\pi}_2^{(2)}$ for maximum likelihood estimation in two exponential models with regression functions of the form \eqref{models-locationscale} is 
\begin{enumerate}
\item[(1)] $\xi^{a,*}=(\xi_1^{a,*}, \xi_2^{a,*}, \mu^{a,*})$ if $\log r \leq 0$ and $\log r \leq u(\tilde{d}^{*,(1)}, \tilde{d}^{*,(2)}) $, where
\begin{equation*}
\xi_1^{a,*} = \begin{Bmatrix} 0 & \tilde{d}^{*,(1)} & d_{max}^{(1)} \\ \frac{1}{3} & \frac{1}{3} & \frac{1}{3} \end{Bmatrix}, \quad \xi_2^{a,*} = \begin{Bmatrix} d_{max}^{(2)} \\ 1 \end{Bmatrix}, \quad \mu^{a,*} = \begin{Bmatrix} 1 & 2 \\ \frac{3}{4} & \frac{1}{4} \end{Bmatrix},
\end{equation*}
\item[(2)] $\xi^{b_1,*}=(\xi_1^{b_1,*}, \xi_2^{b_1,*}, \mu^{b_1,*})$ if $u(\tilde{d}^{*,(1)}, \tilde{d}^{*,(2)}) < \log r \leq 0$, where
\begin{equation*}
\xi_1^{b_1,*} = \begin{Bmatrix} 0 & d_{max}^{(1)} \\ \frac{1}{2} & \frac{1}{2} \end{Bmatrix}, \quad \xi_2^{b_1,*} = \begin{Bmatrix} \tilde{d}^{*,(2)} & d_{max}^{(2)} \\ \frac{1}{2} & \frac{1}{2} \end{Bmatrix}, \quad \mu^{b_1,*} = \begin{Bmatrix} 1 & 2 \\ \frac{1}{2} & \frac{1}{2} \end{Bmatrix},
\end{equation*}
\item[(3)] $\xi^{b_2,*}=(\xi_1^{b_2,*}, \xi_2^{b_2,*}, \mu^{b_2,*})$ if $0 \leq \log r < u(\tilde{d}^{*,(1)}, \tilde{d}^{*,(2)})$, where
\begin{equation*}
\xi_1^{b_2,*} = \begin{Bmatrix} \tilde{d}^{*,(1)} & d_{max}^{(1)} \\ \frac{1}{2} & \frac{1}{2} \end{Bmatrix}, \quad \xi_2^{b_2,*} = \begin{Bmatrix} 0 & d_{max}^{(2)} \\ \frac{1}{2} & \frac{1}{2} \end{Bmatrix}, \quad \mu^{b_2,*} = \begin{Bmatrix} 1 & 2 \\ \frac{1}{2} & \frac{1}{2} \end{Bmatrix},
\end{equation*}
\item[(4)] $\xi^{c,*}=(\xi_1^{c,*}, \xi_2^{c,*}, \mu^{c,*})$ if $\log r \geq 0 $ and $\log(r) > u(\tilde{d}^{*,(1)}, \tilde{d}^{*,(2)})$, where
\begin{equation*}
\xi_1^{c,*} = \begin{Bmatrix} d_{max}^{(1)} \\ 1 \end{Bmatrix}, \quad \xi_2^{c,*} = \begin{Bmatrix} 0 & \tilde{d}^{*,(2)} & d_{max}^{(2)} \\ \frac{1}{3} & \frac{1}{3} & \frac{1}{3} \end{Bmatrix}, \quad \mu^{c,*} = \begin{Bmatrix} 1 & 2 \\ \frac{1}{4} & \frac{3}{4} \end{Bmatrix}.
\end{equation*}
\end{enumerate}
\end{theorem}

\begin{theorem}\label{designs-log}
Let $r=\frac{\sigma_1^2}{\sigma_2^2}$ and $\tilde{d}^{*,(i)}$ be a solution in the interval $(0,d_{max}^{(i)})$ of the equation \eqref{eq-log-ind} using the marginal prior distribution $\tilde{\pi}_2^{(i)}$ on $\Theta_2^{(i)}$. Also define the function
%\begin{align*}
%&u(\tilde{d}^{*,(1)}, \tilde{d}^{*,(2)}) := \int_{\Theta_2^{(1)}} 2 \log \Bigg[ \log\Big(\frac{d_{max}^{(1)}}{\theta_2^{(1)}}+1\Big) \tilde{d}^{*,(1)}(\theta_2^{(1)} + d_{max}^{(1)}) - \log \Big( \frac{\tilde{d}^{*,(1)}}{\theta_2^{(1)}} + 1 \Big) d_{max}^{(1)} (\theta_2^{(1)} + \tilde{d}^{*,(1)})  \Bigg] \\
%&- \frac{2}{\theta_2^{(1)}} \tilde{\pi}_2^{(1)}(d \theta_2^{(1)}) + \int_{\Theta_2^{(2)}} \frac{2}{\theta_2^{(2)}} - 2 \log \Bigg[ \log\Big(\frac{d_{max}^{(2)}}{\theta_2^{(2)}}+1\Big) \tilde{d}^{*,(2)}(\theta_2^{(2)} + d_{max}^{(2)}) - \log \Big( \frac{\tilde{d}^{*,(2)}}{\theta_2^{(2)}} + 1 \Big) d_{max}^{(2)} \\
%&(\theta_2^{(2)} + \tilde{d}^{*,(2)})  \Bigg] \tilde{\pi}_2^{(2)}(d \theta_2^{(2)}).
%\end{align*}
\begin{align*}
&u(\tilde{d}^{*,(1)}, \tilde{d}^{*,(2)}) \\
& := 2\sum_{i=1}^{2} (-1)^{i+1} \int_{\Theta_2^{(i)}} \log \Bigg[\Big( \tfrac{\tilde{d}^{*,(i)}}{\tilde{d}^{*,(i)} + \theta_2^{(i)}}\Big) \Big(g(d_{max}^{(i)}, \theta_2^{(i)}) - g(\tilde{d}^{*,(i)}, \theta_2^{(i)}) \Big) \Bigg] \tilde{\pi}_2^{(i)}(d \theta_2^{(i)}) 
\end{align*}
where $g(d, \theta_2) = \tfrac{d+ \theta_2}{d} \log(\tfrac{d}{\theta_2}+1)$.
The Bayesian $D$-optimal design in the class $\Xi_4^2$ with respect to the prior $\pi= \tilde{\pi}_1 \times \tilde{\pi}_2 \times \tilde{\pi}_2^{(1)} \times \tilde{\pi}_2^{(2)}$ for maximum likelihood estimation in two linear-in-log models with regression functions of the form \eqref{models-locationscale} is 
\begin{enumerate}
\item[(1)] $\xi^{a,*}=(\xi_1^{a,*}, \xi_2^{a,*}, \mu^{a,*})$ if $\log r \leq 0$ and $\log r \leq u(\tilde{d}^{*,(1)}, \tilde{d}^{*,(2)}) $, where
\begin{equation*}
\xi_1^{a,*} = \begin{Bmatrix} 0 & \tilde{d}^{*,(1)} & d_{max}^{(1)} \\ \frac{1}{3} & \frac{1}{3} & \frac{1}{3} \end{Bmatrix}, \quad \xi_2^{a,*} = \begin{Bmatrix} d_{max}^{(2)} \\ 1 \end{Bmatrix}, \quad \mu^{a,*} = \begin{Bmatrix} 1 & 2 \\ \frac{3}{4} & \frac{1}{4} \end{Bmatrix},
\end{equation*}
\item[(2)] $\xi^{b_1,*}=(\xi_1^{b_1,*}, \xi_2^{b_1,*}, \mu^{b_1,*})$ if $u(\tilde{d}^{*,(1)}, \tilde{d}^{*,(2)}) < \log r \leq 0$, where
\begin{equation*}
\xi_1^{b_1,*} = \begin{Bmatrix} 0 & d_{max}^{(1)} \\ \frac{1}{2} & \frac{1}{2} \end{Bmatrix}, \quad \xi_2^{b_1,*} = \begin{Bmatrix} \tilde{d}^{*,(2)} & d_{max}^{(2)} \\ \frac{1}{2} & \frac{1}{2} \end{Bmatrix}, \quad \mu^{b_1,*} = \begin{Bmatrix} 1 & 2 \\ \frac{1}{2} & \frac{1}{2} \end{Bmatrix},
\end{equation*}
\item[(3)] $\xi^{b_2,*}=(\xi_1^{b_2,*}, \xi_2^{b_2,*}, \mu^{b_2,*})$ if $0 < \log r < u(\tilde{d}^{*,(1)}, \tilde{d}^{*,(2)})$, where
\begin{equation*}
\xi_1^{b_2,*} = \begin{Bmatrix} \tilde{d}^{*,(1)} & d_{max}^{(1)} \\ \frac{1}{2} & \frac{1}{2} \end{Bmatrix}, \quad \xi_2^{b_2,*} = \begin{Bmatrix} 0 & d_{max}^{(2)} \\ \frac{1}{2} & \frac{1}{2} \end{Bmatrix}, \quad \mu^{b_2,*} = \begin{Bmatrix} 1 & 2 \\ \frac{1}{2} & \frac{1}{2} \end{Bmatrix},
\end{equation*}
\item[(4)] $\xi^{c,*}=(\xi_1^{c,*}, \xi_2^{c,*}, \mu^{c,*})$ if $\log r \geq 0 $ and $\log(r) > u(\tilde{d}^{*,(1)}, \tilde{d}^{*,(2)})$, where
\begin{equation*}
\xi_1^{c,*} = \begin{Bmatrix} d_{max}^{(1)} \\ 1 \end{Bmatrix}, \quad \xi_2^{c,*} = \begin{Bmatrix} 0 & \tilde{d}^{*,(2)} & d_{max}^{(2)} \\ \frac{1}{3} & \frac{1}{3} & \frac{1}{3} \end{Bmatrix}, \quad \mu^{c,*} = \begin{Bmatrix} 1 & 2 \\ \frac{1}{4} & \frac{3}{4} \end{Bmatrix}.
\end{equation*}
\end{enumerate}
\end{theorem}

\section{Example}\label{secex}

In this section we assess the Bayesian $D$-optimal saturated designs characterised analytically in Section 3, via a numerical investigation of their performance in a situation where these are not Bayesian $D$-optimal amongst all designs.

In particular, we consider the case of $M=2$ different administration frequency groups where the Emax regression function is used to model the dose-response relationship for both groups. This choice of parametric model is frequent in dose-response studies see, for example, \citet{extra1} and \citet{extra2} for recent references. It is also reasonable to assume that the placebo effect as well as the maximum effect that can be achieved by the drug under investigation is the same between groups. Therefore, the two Emax models are assumed to have common location and scale parameters which corresponds to the case considered in Section 3.3. Furthermore, for illustration purposes, the design spaces $\mathcal{X}_1$ and $\mathcal{X}_2$ are assumed to coincide with $\mathcal{X}_1= \mathcal{X}_2= [0, 1]$ and also the variances in both groups are assumed to be the same with $\sigma^2_1= \sigma^2_2= 1$.

 In what follows, the prior $\pi= \tilde\pi_1 \times \tilde\pi_2 \times \tilde\pi^{(1)}_2 \times \tilde\pi^{(2)}_2 $ on the parameter space $\Theta= \Theta_1 \times \Theta^{(1)}_2 \times \Theta^{(2)}_2$ is used. The marginal priors $\tilde\pi_1$ and $\tilde\pi_2$  are set to the dirac measures $\delta_{\theta_1}$ and $\delta_{\theta_2}$ with $(\theta_1, \theta_2)^T=(0, 1)^T$, respectively, since the Emax model is linear in the location and scale parameter and varying these parameters does not affect the information matrix.  For the priors  $\tilde\pi^{(1)}_2$ and $\tilde\pi^{(2)}_2$ we use uniform distributions on the sets $\{0.20, 0.275, 0.35, 0.425, 0.50\} $ and $\{0.60, 0.675, 0.75, 0.825, 0.90\}$, respectively and thus, we have a total of $P=5 \cdot 5= 25$ different and equally probable parameter combinations. This corresponds to the case of no preference for specific parameter combinations and thus under the concept of uniform distributions there is no need for a prior to be specified, a step which is often difficult in practice.

The scenario described above corresponds to the first case of Theorem \ref{designs-emax} according to which we calculate the Bayesian $D$-optimal design $\xi^{*}= (\xi_1^{*} , \xi_2^{*}, \mu^{*} )$ in the class $\Xi_4^2$ of saturated designs with respect to the prior $\pi$ which is given by 
\begin{equation}\label{saturated-numerics}
\xi_1^{*} = \begin{Bmatrix} 0.00 & 0.1984207 & 1.00\\ \frac{1}{3} & \frac{1}{3} & \frac{1}{3}\end{Bmatrix}, \quad \xi_2^{*} = \begin{Bmatrix} 0.742427 \\ 1 \end{Bmatrix}, \quad \mu^{*} = \begin{Bmatrix} 1 & 2 \\ \frac{3}{4} & \frac{1}{4} \end{Bmatrix}.
\end{equation}

\begin{figure}[h!]
	\centering
  \includegraphics[width=0.5\textwidth]{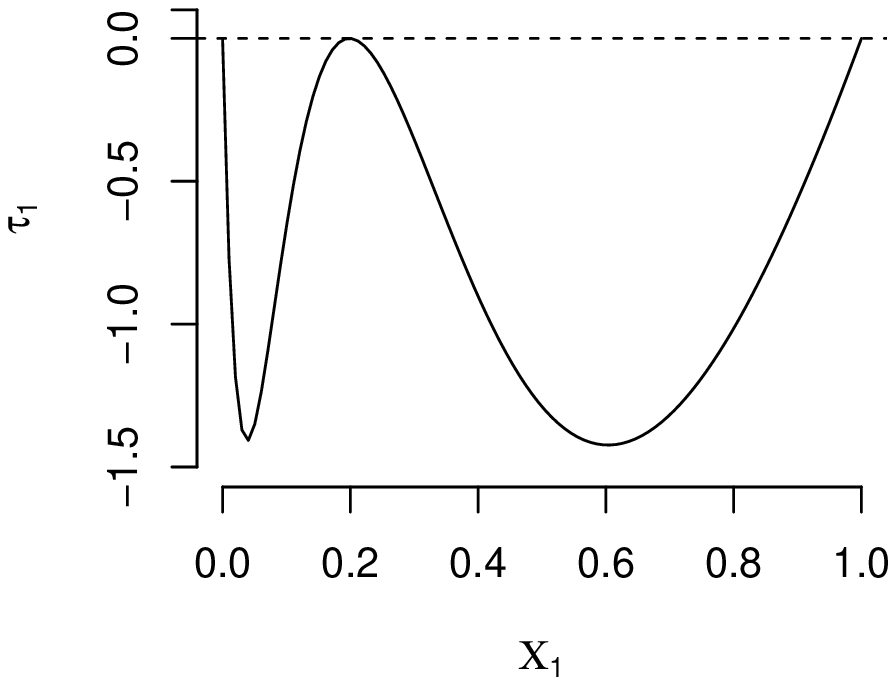}~
  \includegraphics[width=0.5\textwidth]{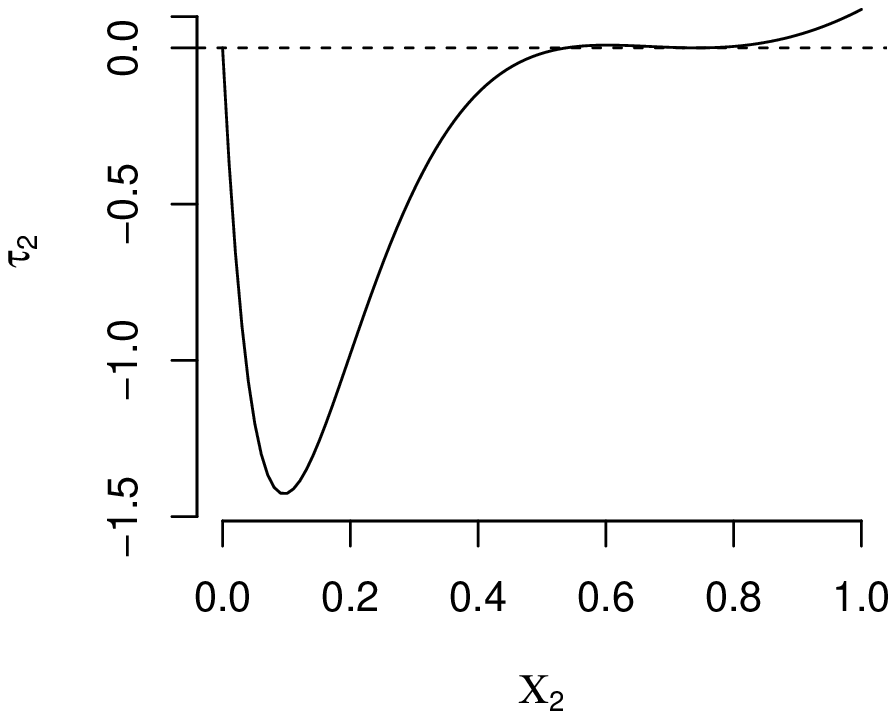}
	\caption{\it Illustration of Theorem \ref{get} for the Bayesian $D$-optimal design $\xi^*$ in the class $\Xi_4^2$. The figures show the functions $\tau_1$ and $\tau_2$ defined by the left-hand side of inequality \eqref{tau_get}, respectively. Left panel: $\tau_1$. Right panel: $\tau_2$.  }
	\label{fig1}
\end{figure}

Checking the inequalities \eqref{tau_get} of the Equivalence Theorem \ref{get} reveals that the design $\xi^*$ is not Bayesian $D$-optimal amongst all designs since the inequality is not satisfied for the second administration frequency group (see Figure \ref{fig1}).

Nevertheless, we further investigate the performance of the design $\xi^*$ given in \eqref{saturated-numerics} since if it performs well then, as it is discussed in Section 3.1, it is preferred to the design that is Bayesian $D$-optimal amongst all designs. The reason is that $\xi^*$ has minimum number of support points, that is, it requires fewer doses to be utilised in the two administration groups and thus it results in lower costs for the dose-response study without loosing much information. We also note that finding the design that is Bayesian $D$-optimal amongst all possible designs is computationally intensive even for the case of only two administration frequency groups and to the best of our knowledge no analytical characterisations of these designs, which would facilitate the evaluation, exist in the literature.

For the calculation of the design which is Bayesian $D$-optimal amongst all designs we used the particle swarm optimisation algorithm (see \cite{Clerc2006} for details) and the resulting candidate design $\xi^{*,B}= (\xi_1^{*,B} , \xi_2^{*,B}, \mu^{*,B} )$ is given by

\begin{equation}\label{bayesian-numerics}
\begin{split}
\xi_1^{*,B} &= \begin{Bmatrix} 0.19982 & 1.00 \\ 0.50148  & 0.49852\end{Bmatrix}, \quad \xi_2^{*,B} = \begin{Bmatrix} 0.00 & 0.56386 & 1.00 \\ 0.48649 &  0.26260 & 0.25091 \end{Bmatrix},\\
\mu^{*,B} &= \begin{Bmatrix} 1 & 2 \\ 0.48691 & 0.51309 \end{Bmatrix}.
\end{split}
\end{equation}

\begin{figure}[h!]
	\centering
  \includegraphics[width=0.5\textwidth]{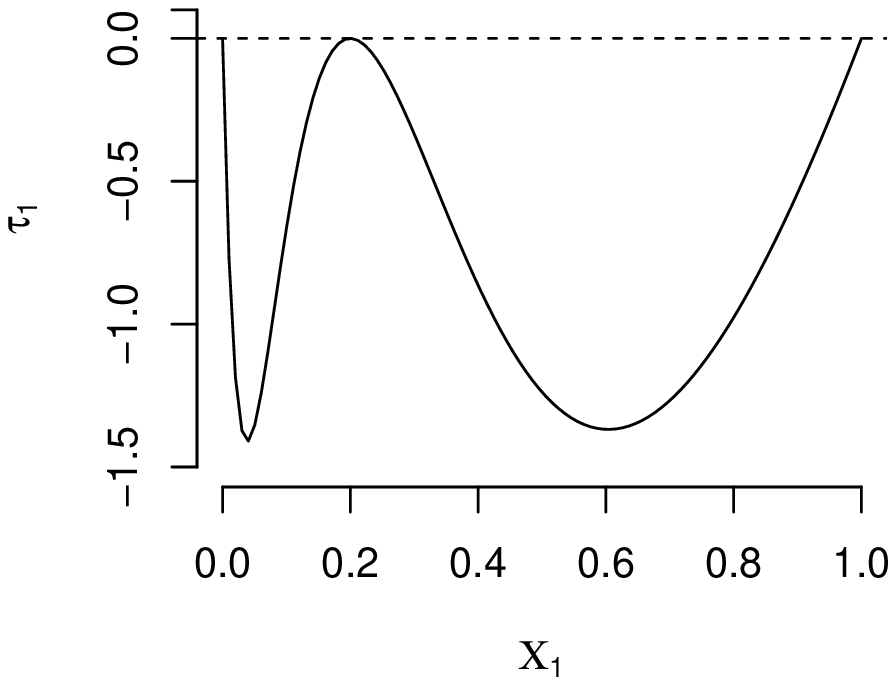}~
  \includegraphics[width=0.5\textwidth]{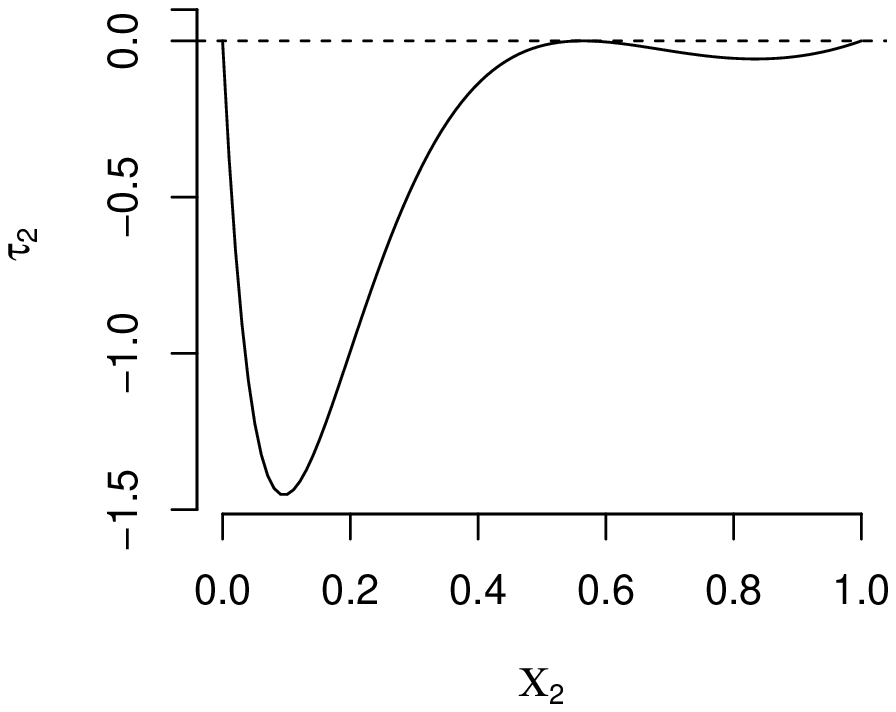}
	\caption{\it Illustration of Theorem \ref{get} for the Bayesian $D$-optimal design $\xi^{*,B}$. The figures show the functions $\tau_1$ and $\tau_2$ defined by the left-hand side of inequality \eqref{tau_get}, respectively. Left panel: $\tau_1$. Right panel: $\tau_2$.  }
	\label{fig2}
\end{figure}

As shown in Figure \ref{fig2} both inequalities \eqref{tau_get} of the Equivalence Theorem \ref{get} are satisfied for this candidate design and therefore, $\xi^{*,B}$ given in \eqref{bayesian-numerics} is in fact Bayesian $D$-optimal amongst all designs.

We now assess the performance of a given design $\xi$ (for a specific parameter vector $\boldsymbol{\theta}$) via the $D$-efficiency defined as 
\begin{equation}\label{eff}
\mbox{eff}(\xi,\boldsymbol{\theta})= \frac{\mbox{det}(M(\xi, \boldsymbol{\theta}))^{1/4}}{\mbox{det}(M(\xi^*(\boldsymbol{\theta}), \boldsymbol{\theta}))^{1/4}} \in [0,1],
\end{equation} 
where $\xi^*(\boldsymbol{\theta})$ is the locally $D$-optimal design with respect to the specific parameter vector $\boldsymbol{\theta}$. The closer the $D$-efficiency of $\xi$ is to 1, the fewer replicates are required for the design $\xi$ to be able to estimate the parameters with the same precision as $\xi^*(\boldsymbol{\theta})$.

Table \ref{tab2} presents the $D$-efficiencies of  $\xi^*$ and $\xi^{*,B}$, given in \eqref{saturated-numerics} and \eqref{bayesian-numerics} respectively, for the various parameter combinations resulting from the choice of prior $\pi$. For example, for the parameter vector $\boldsymbol{\theta}=(0, 1, 0.2, 0.6)$ the $D$-efficiencies of $\xi^*$ and $\xi^{*,B}$ are $97.52\%$ and $97.31\%$, respectively.

\begin{table}[h!]
\centering
\begin{tabular}{|l|rr||rrrrr|}
  \hline
  & & &  \multicolumn{5}{c|}{$\theta_2^{(1)}$} \\
  & &  & 0.20 & 0.275 & 0.35 &  0.425 &  0.50\\ 
  \hline
  \hline
  \multirow{5}{*}{ $\mbox{eff}(\xi^{*},(0, 1, \theta_2^{(1)}, \theta_2^{(2)}))$} &   \multirow{5}{*}{ $\theta_2^{(2)}$} 	& 0.6 & 97.52 & 99.20 & 98.68 & 96.94 & 94.25 \\
												 %\cline{3-8}
												 & & 0.675 & 97.96 & 99.65 & 99.50 & 98.26 & 96.23 \\
												 												% \cline{3-8}
												 & & 0.75 & 98.07 & 99.76 & 99.81 & 98.91 & 97.37\\
												 												% \cline{3-8}
												 & & 0.825 & 97.94 & 99.62 & 99.77 & 99.10 & 97.90 \\
												 												 %\cline{3-8}
												 & & 0.90 &97.62 & 99.30 & 99.51 & 98.97 & 98.02\\
 \hline
 \hline
\multirow{5}{*}{ $\mbox{eff}(\xi^{*,B},(0, 1, \theta_2^{(1)}, \theta_2^{(2)}))$} &   \multirow{5}{*}{ $\theta_2^{(2)}$} 	& 0.6 & 97.31 & 99.39 & 99.41 & 98.38 & 96.59 \\
												%\cline{3-8}
												 & & 0.675 & 97.58 &99.59 & 99.88 & 99.22 & 97.95 \\
												 % \cline{3-8}
												 & & 0.75 & 97.58 & 99.53 & 99.94 & 99.52 & 98.60\\
												 %\cline{3-8}
												 & & 0.825 &97.40 & 99.30 & 99.74 & 99.46 & 98.78 \\
												 %\cline{3-8}
												 & & 0.90 &97.07  & 98.93 &  99.38 &  99.16 & 98.63\\
\hline
\end{tabular}
\caption{\it The efficiencies (in \%) of the Bayesian $D$-optimal design $\xi^*$ in the class $\Xi_4^2$ and of the Bayesian $D$-optimal design $\xi^{*, B}$ for the $P=25$ different parameter combinations.}
\label{tab2}
\end{table}

Note that the $D$-efficiencies of both designs $\xi^*$ and $\xi^{*, B}$ depicted in Table \ref{tab2} are between $94\%$ and $99\%$. Consequently, using either one of these two designs results in parameter estimation with accuracy very close to the one the locally $D$-optimal design $\xi^*(\boldsymbol{\theta})$ would have achieved. Moreover, for every parameter combination the $D$-efficiency of the design $\xi^*$ which is Bayesian $D$-optimal only in the class of saturated designs is close to the corresponding $D$-efficiency of the design $\xi^{*, B}$ which is Bayesian $D$-optimal amongst all designs. We therefore, conclude that our analytical characterisations of Bayesian $D$-optimal saturated designs can be used without detriment thus reducing both the numerical effort for design search as well as the cost of the dose-response study.

\bigskip

{\bf Acknowledgements}\\
This work has been supported in part by the
Collaborative Research Center "Statistical modeling of nonlinear
dynamic processes" (SFB 823, Project C2) of the German Research Foundation (DFG). Schorning was partially supported  by a grant from the National Institute Of General Medical Sciences of the National
Institutes of Health under Award Number R01GM107639. The content is solely the responsibility of the author and does not necessarily
 represent the official views of the National Institutes of Health).

\appendix{}
\section{Appendix}

\textbf{\emph{Proof of Lemma \ref{equal-weights-ind}}}:
\begin{proof}
Let $\tilde{\xi} = \{ d_1, \ldots, d_{\tilde{m}}; \omega_1, \ldots, \omega_{\tilde{m}} \}$ be any saturated design on $\mathcal{X}=[0, d_{max}]$. Also let $\tilde{X}$ be the $\tilde{m} \times \tilde{m}$ matrix with $j$th column given by $\tilde{h}(d_j, \tilde{\boldsymbol{\theta}})$, $j=1, \ldots, \tilde{m}$ and $\tilde{W} = \text{diag}(\omega_1, \ldots, \omega_{\tilde{m}})$. Using this notation the information matrix defined in  \eqref{info-ind} and \eqref{deriv-ind} becomes 
\begin{equation*}
M_{\text{ind}}(\tilde{\xi}, \tilde{\boldsymbol{\theta}}) = \tilde{X} \tilde{W} \tilde{X}^T,
\end{equation*}
and thus its determinant is given by
\begin{equation*}
\det \left\{ M_{\text{ind}}(\tilde{\xi}, \tilde{\boldsymbol{\theta}}) \right\}= [\det \{\tilde{X}\} ]^2 \det \{\tilde{W}\}.
\end{equation*}
Hence the Bayesian $D$-optimality criterion given in \eqref{criterion-ind} becomes
\begin{equation*}
\Phi_{\tilde{\pi}}(\tilde{\xi}) = \int_{\tilde{\Theta}} \left( 2 \log |\det \{ \tilde{X} \}| + \sum_{j=1}^{\tilde{m}} \log \omega_j \right) \tilde{\pi}(d \tilde{\boldsymbol{\theta}}).
\end{equation*}
Maximising the above expression with respect to the weights $\omega_1, \ldots, \omega_{\tilde{m}}$ under the condition $\sum_{j=1}^{\tilde{m}} \omega_j = 1$ gives $\omega_j = \frac{1}{\tilde{m}}$ for all $j=1, \ldots, \tilde{m}$ which proves the assertion. 

\end{proof}

\textbf{\emph{Proof of Theorem \ref{designs-ind}}}:
\begin{proof}
From Lemma \ref{equal-weights-ind} we know that a Bayesian $D$-optimal saturated design has equal weights. Therefore, let $\tilde{\xi}=\{ d_1, d_2, d_3; 1/3, 1/3, 1/3 \}$ be an equally weighted three-point design where $0 \leq d_1 < d_2 < d_3 \leq d_{max}$. 

Theorem 4.1 in \citet{detkisbevbre2010} shows that for any of the models \eqref{emax}, \eqref{exp} and \eqref{log} the locally $D$-optimal design is supported at exactly three points two of which are the end-points of the design space. Therefore, for any of these models the determinant of the information matrix \eqref{info-ind} of the equally weighted three-point design $\tilde{\xi}$ is decreasing with $d_1$ and increasing $d_3$. 

As a result, the Bayesian $D$-optimality criterion function given in \eqref{criterion-ind} is also decreasing with $d_1$ and increasing with $d_3$ and thus maximised at $d_1^*=0$ and $d_3^*=d_{max}$. That is, for any of the models \eqref{emax}, \eqref{exp} and \eqref{log} a Bayesian $D$-optimal saturated design is always supported at the end-points of the design space. The non-trivial support point can be found by solving the equation
\begin{equation*}
\frac{\partial}{\partial d_2} \Phi_{\tilde{\pi}}(\tilde{\xi}) | _{d_1=0,d_3=d_{max}} = 0 \iff \int_{\tilde{\Theta}_2} \frac{\partial}{\partial d_2} \left( \log \det\{M_{\text{ind}}(\tilde{\xi}, \tilde{\boldsymbol{\theta}})\} \right) | _{d_1=0,d_3=d_{max}} \,\tilde{\pi}_2(d \theta_2) =0,
\end{equation*}
for $d_2 \in (0,d_{max})$. Using the appropriate information matrix for each of the three models under consideration yields the equations of Theorem \ref{designs-ind}. 

\end{proof}

\textbf{\emph{Proof of Theorem \ref{get}}}:\\
\begin{proof}
For a fixed $\boldsymbol{\theta}$ consider the set 
\begin{equation*}
\mathcal{M}_{\boldsymbol{\theta}} = \left\{ M(\xi_{\pi},\boldsymbol{\theta}) | \xi_{\pi} = (\xi_1, \ldots, \xi_M,\mu) \in \Xi_1(\mathcal{X}_1)  \times \ldots \times \Xi_M(\mathcal{X}_M) \times \Xi(\{1,\ldots,M \}) \right\} ,
\end{equation*}
where $\Xi_1(\mathcal{X}_1)  \times \ldots \times \Xi_M(\mathcal{X}_M) \times \Xi(\{1,\ldots,M \})$ denotes the class of all designs on $\mathcal{X}_1 \times \ldots \times \mathcal{X}_M \times  \{1,\ldots,M \}$, and note that $\mathcal{M}_{\boldsymbol{\theta}}$  is the convex hull of
\begin{equation*}
\mathcal{D}_{\boldsymbol{\theta}}  = \left\{ M^{(i)}(\delta_d, \boldsymbol{\theta}) = h_i(d,\boldsymbol{\theta}) h_i^T(d,\boldsymbol{\theta}) | i \in \{ 1, \ldots, M \}, d \in \mathcal{X}_i \right\}.
\end{equation*}

Using Theorem 3.3 in \citet{dette2007}, a design $\xi^*_{\pi} = (\xi_1^*, \ldots, \xi_M^*,\mu^*)$ is Bayesian $D$-optimal, with respect to the prior $\pi$, in the class $\Xi_1(\mathcal{X}_1)  \times \ldots \times \Xi_M(\mathcal{X}_M) \times \Xi(\{1,\ldots,M \})$ if and only if the inequality
\begin{equation*}
\int_{\Theta} \text{tr} \{ (M(\xi_{\pi}^*,\boldsymbol{\theta}))^{-1} E_{\boldsymbol{\theta}} \} \pi(d\boldsymbol{\theta}) \leq m,
\end{equation*}
holds for all matrices $E_{\boldsymbol{\theta}} \in \mathcal{M}_{\boldsymbol{\theta}}$.

Since $\mathcal{M}_{\boldsymbol{\theta}} $ is the convex hull of $\mathcal{D}_{\boldsymbol{\theta}}$, it is sufficient to prove inequality for all $E_{\boldsymbol{\theta}} \in \mathcal{D}_{\boldsymbol{\theta}}$. Therefore, $\xi^*_{\pi}$ is Bayesian $D$-optimal, with respect to the prior $\pi$, in the class $\Xi_1(\mathcal{X}_1)  \times \ldots \times \Xi_M(\mathcal{X}_M) \times \Xi(\{1,\ldots,M \})$ if and only if the inequality
\begin{equation*}
\int_{\Theta} \text{tr}\{ (M(\xi_{\pi}^*,\boldsymbol{\theta}))^{-1} h_i(d,\boldsymbol{\theta}) h_i^T(d,\boldsymbol{\theta}) \} \pi(d\boldsymbol{\theta}) \leq m,
\end{equation*}
holds for all $d \in \mathcal{X}_i$ and $i=1, \ldots, M$. Rewriting the above gives the inequalities of Theorem \ref{get}.
\end{proof}

\textbf{\emph{Proof of Lemma \ref{equal-weights}}}:
\begin{proof}
The information matrix of the design $\xi_{\pi}$ can be rewritten as 
\begin{equation*}
M(\xi_{\pi}, \boldsymbol{\theta}) = X W X^T,
\end{equation*}
where 
\begin{equation*}
X = \left( h_1(d_1^{(1)}), \ldots, h_1(d_{m_1}^{(1)}), \ldots, h_M(d_{1}^{(M)}), \ldots, h_M(d_{m_M}^{(M)})  \right) \in \mathbb{R}^{m \times m},
\end{equation*}
and 
\begin{equation*}
W = \text{diag}\left( \lambda_1 \omega_{11}, \ldots, \lambda_1 \omega_{1m_1}, \ldots, \lambda_M \omega_{M1}, \ldots, \lambda_M \omega_{Mm_M} \right) \in \mathbb{R}^{m \times m}.
\end{equation*}

Then the Bayesian $D$-optimality criterion function given in \eqref{criterion} becomes 
\begin{equation*}
\int_{\Theta} \log \det\{ M(\xi_{\pi}, \boldsymbol{\theta}) \} \pi(d \boldsymbol{\theta}) = \int_{\Theta} \left( 2 \log |\det \{X\}| + \log \det \{W\} \right) \pi(d \boldsymbol{\theta}).
\end{equation*}
Maximising 
\begin{equation*}
\log \det\{W\} = \sum_{i=1}^{M} \log \lambda_i \sum_{j=1}^{m_i} \log \omega_{ij},
\end{equation*}
under the conditions $\sum_{i=1}^{M} \lambda_i =1$, $\sum_{j=1}^{m_i} \omega_{ij} =1$ and $\sum_{i=1}^{M} m_i = m$ gives
\begin{equation*}
\lambda_i = \frac{m_i}{m}, \qquad \omega_{ij} = \frac{1}{m_i}, (i=1, \ldots, M; j=1,\ldots, m_i),
\end{equation*}
which proves the assertion. 

\end{proof}

\textbf{\emph{Proof of Theorem \ref{designs-location-general}}}
\begin{proof}
For the sake of transparency we restrict ourselves to the case $M=2$ such that $m=2q+1$. \\
By an application of Lemma \ref{equal-weights} we obtain the following two candidates $\xi^a=(\xi^a_1, \xi^a_2, \mu^a)$ and $\xi^b=(\xi^b_1, \xi^b_2, \mu^b)$ for a Bayesan $D$-optimal design in the class $\Xi_{2q+1}^2$ with
\begin{eqnarray*}
\xi^a_1 &=& \begin{Bmatrix} d^{(1)}_0 & d^{(1)}_{1} & \ldots & d^{(1)}_q  \\ \frac{1}{q+1} &  \frac{1}{q+1} & \ldots & \frac{1}{q+1} \end{Bmatrix}, \quad
\xi^a_2 = \begin{Bmatrix} d^{(2)}_1 & \ldots & d^{(2)}_q  \\ \frac{1}{q} & \ldots & \frac{1}{q}  \end{Bmatrix}, \quad
\mu^a =   \begin{Bmatrix} 1 & 2 \\ \frac{q+1}{2q+1} & \frac{q}{2q+1} \end{Bmatrix}, \\
\xi^b_1 &=& \begin{Bmatrix} d^{(1)}_1 & \ldots & d^{(1)}_q  \\ \frac{1}{q} & \ldots & \frac{1}{q}  \end{Bmatrix}, \quad
\xi^b_2 = \begin{Bmatrix} d^{(2)}_0 &d^{(2)}_{1} & \ldots & d^{(2)}_{q}  \\ \frac{1}{q+1} & \frac{1}{q+1} & \ldots  & \frac{1}{q+1} \end{Bmatrix} , \quad
\mu^b =  \begin{Bmatrix} 1 & 2 \\ \frac{q}{2q+1} & \frac{q+1}{2q+1} \end{Bmatrix}.
\end{eqnarray*}
We now evaluate the criterion values $\Phi_{\tilde{\pi}}(\xi^{a})$ and $\Phi_{\tilde{\pi}}(\xi^{a})$ and maximise them with respect to the support points of the design $\xi^a$ and $\xi^b$, respectively. \\
For the design $\xi^a$, the criterion value $\Phi_{\tilde{\pi}}(\xi^{a})$ can be rewritten by
\begin{equation}\label{phixia}
\begin{split}
\Phi_{\tilde{\pi}}(\xi^{a}) &= \int_{\Theta^{(1)}_2} (q+1)\log\tfrac{1}{\sigma^2_1} + (q+1)\log\tfrac{1}{q+1}  + 2\log|\mbox{det}\big\{X_1(d^{(1)}_0, d^{(1)}_1,   \ldots , d^{(1)}_{q}, \theta^{(1)}_2 )\big\}|\tilde{\pi}^{(1)}_2(d\theta_2^{(1)}) \\
&+ \int_{\Theta^{(2)}_2} q \log\tfrac{1}{\sigma^2_2} + q \log\tfrac{1}{q} +  2\log|\mbox{det}\big\{X_2(d^{(2)}_1 , \ldots , d^{(2)}_q,\theta^{(2)}_2 )\big\}|\tilde{\pi}^{(2)}_2(d\theta_2^{(2)}) +c(q),
\end{split}
\end{equation}
where the $(q+1)\times(q+1)$-dimensional matrix $X_1(d_0 , d_1, \ldots , d_q, \theta_2 )$ is given by
\begin{equation*}
X_1(d_0 , d_1, \ldots , d_q, \theta_2 )= \begin{pmatrix} 1 &1 & \ldots  & 1 \\ \eta_0(d_0, \theta_2) &  \eta_0(d_{1},  \theta_2) & \ldots & \eta_0(d_q, \theta_2) \end{pmatrix},
\end{equation*}
the $q \times q$-dimensional matrix $X_2(d_1 , \ldots , d_q, \theta_2 )$ is given by
\begin{equation*}
X_2(d_1 , \ldots , d_q, \theta_2 )= \begin{pmatrix} \eta_0(d_1, \theta_2) & \ldots & \eta_0(d_q, \theta_2) \end{pmatrix},
\end{equation*}
the constant $c(q)$ only depends on the number of the non-common parameters $q$. 

The first term of the sum in \eqref{phixia} concides with Bayesian $D$-criterion for the individual model of the form \eqref{models-location} with respect to the prior ${\tilde{\pi}^{(1)}_2}$ and is therefore maximised by the corresponding Bayesian $D$-optimal design $\tilde\xi^*_{\tilde{\pi}^{(1)}_2}$ given by \eqref{individualdesign}. 
Because of the conditions in \eqref{cond32} the second term of the sum in \eqref{phixia} can be rewritten by 
\begin{equation*}
\begin{split}
&\int_{\Theta^{(2)}_2} q \log\tfrac{1}{\sigma^2_2} + q \log\tfrac{1}{q} +  2\log | \mbox{det}\big\{X_2(d^{(2)}_1 , \ldots , d^{(2)}_q,\theta^{(2)}_2 )\big\} | \tilde{\pi}^{(2)}_2(d\theta_2^{(2)}) \\
= &\int_{\Theta^{(2)}_2} q \log\tfrac{1}{\sigma^2_2} + q \log\tfrac{1}{q} +  2\log | \mbox{det}\big\{X_1(0, d^{(2)}_1 , \ldots , d^{(2)}_q,\theta^{(2)}_2 )\big\}|\tilde{\pi}^{(2)}_2(d\theta_2^{(2)}),
\end{split}
\end{equation*}
where the latter expression again corresponds to the Bayesian $D$-criterion for the individual model of the form \eqref{models-location} with respect to the prior ${\tilde{\pi}^{(2)}_2}$ under the additional constraint that the smallest support point $d^{(2)}_0=0$ is fixed to zero. Maximising this expression with respect to the remaining $q$ support points $d^{(2)}_1, \ldots, d^{(2)}_q$ we receive that $\xi_2^a = \xi^{*}_2$ where $\xi^{*}_2$ is defined by \eqref{32alldes}.

A similar calculation for the criterion value $\Phi_{\tilde{\pi}}(\xi^{b})$ results in the maximising design $\xi^b$ where $\xi_1^b= \xi_1^{*}$ and $\xi_2^b =  \tilde\xi^*_{\tilde{\pi}^{(2)}_2}$ which are defined in \eqref{32alldes} and \eqref{individualdesign}, respectively. \\
Taking into account that $\sigma^2_1 \leq \sigma_2^2$ a comparison of the criterion values evaluated in $\xi^a$ and $\xi^b$ results in Theorem \ref{designs-location-general}. 
 \end{proof}

\textbf{\emph{Proof of Theorem \ref{designs-emax}}}:
\begin{proof}
Let $\xi_{\pi}=(\xi_1,\xi_2,\mu)$ be an arbitrary design in the class $\Xi_4^2$ of saturated designs on $\mathcal{X}_1 \times \mathcal{X}_2 \times \{1,2\}$. The information matrix of the design $\xi_{\pi}$ is given by 
\begin{equation*}
M(\xi_{\pi}, \boldsymbol{\theta}) = \lambda \int_0^{d_{max}^{(1)}} h_1(d, \boldsymbol{\theta}) h_1^T(d, \boldsymbol{\theta}) \,d \xi_1(\boldsymbol{\theta}) + (1-\lambda) \int_0^{d_{max}^{(2)}} h_2(d, \boldsymbol{\theta}) h_2^T(d, \boldsymbol{\theta}) \,d \xi_2(\boldsymbol{\theta}) ,
\end{equation*}
where for Emax models 
\begin{equation*}
h_1(d, \boldsymbol{\theta}) = \frac{1}{\sigma_1} \left( 1, \frac{d}{d+\theta_2^{(1)}}, \frac{d}{(d+\theta_2^{(1)})^2}, 0 \right)^T, \quad h_2(d, \boldsymbol{\theta}) = \frac{1}{\sigma_2} \left( 1, \frac{d}{d+\theta_2^{(2)}}, 0, \frac{d}{(d+\theta_2^{(2)})^2} \right)^T .
\end{equation*}

Let $\pi=\tilde{\pi}_1 \times \tilde{\pi}_2 \times \tilde{\pi}_2^{(1)} \times \tilde{\pi}_2^{(2)}$ be a prior distribution on the parameter space $\Theta = \Theta_1 \times \Theta_2 \times \Theta_2^{(1)} \times \Theta_2^{(2)}$. By an application of Lemma \ref{equal-weights} we obtain the following candidate designs for the Bayesian $D$-optimal design in the class $\Xi_4^2$.
\begin{equation*}
\xi_1^a = \begin{Bmatrix} d_1^{(1)} & d_2^{(1)} & d_3^{(1)} \\ \frac{1}{3} & \frac{1}{3} & \frac{1}{3} \end{Bmatrix}, \quad \xi_2^a = \begin{Bmatrix} d_1^{(2)} \\ 1 \end{Bmatrix}, \quad \mu^a = \begin{Bmatrix} 1 & 2 \\ \frac{3}{4} & \frac{1}{4} \end{Bmatrix},
\end{equation*}
\begin{equation*}
\xi_1^b = \begin{Bmatrix} d_1^{(1)} & d_2^{(1)} \\ \frac{1}{2} & \frac{1}{2} \end{Bmatrix}, \quad \xi_2^b = \begin{Bmatrix} d_1^{(2)} & d_2^{(2)} \\ \frac{1}{2} & \frac{1}{2} \end{Bmatrix}, \quad \mu^b = \begin{Bmatrix} 1 & 2 \\ \frac{1}{2} & \frac{1}{2} \end{Bmatrix},
\end{equation*}
\begin{equation*}
\xi_1^c = \begin{Bmatrix} d_1^{(1)} \\ 1 \end{Bmatrix}, \quad \xi_2^c = \begin{Bmatrix} d_1^{(2)} & d_2^{(2)} & d_3^{(2)} \\ \frac{1}{3} & \frac{1}{3} & \frac{1}{3} \end{Bmatrix}, \quad \mu^c = \begin{Bmatrix} 1 & 2 \\ \frac{1}{4} & \frac{3}{4} \end{Bmatrix}.
\end{equation*}
We now evaluate the Bayesian $D$-optimality criterion function given in \eqref{criterion} for each of these candidate designs and maximise it with respect to the support points.

For the design $\xi_{\pi}^a=(\xi_1^a, \xi_2^a, \mu^a)$ the criterion function becomes
\begin{align*}
\Phi_{\pi}(\xi_{\pi}^a) &= \int_{\Theta_2 \times \Theta_2^{(1)}} \log \left[ \det \left\{ M_{\text{ind}}(\xi_1^a, \theta_2, \theta_2^{(1)}) \right\} \right] (\tilde{\pi}_2 \times \tilde{\pi}_2^{(1)})(d \theta_2 d \theta_2^{(1)}) \\
&+ \int_{\Theta_2 \times \Theta_2^{(2)}} \log \left[ \left( \frac{1}{4} \right)^4 3^3 \frac{1}{\sigma_2^2} \theta_2^2 \left( \frac{d_1^{(2)}}{(d_1^{(2)}+\theta_2^{(2)})^2} \right)^2 \right] (\tilde{\pi}_2 \times \tilde{\pi}_2^{(2)})(d \theta_2 d \theta_2^{(2)}).
\end{align*}
From Theorem \ref{designs-ind} we have that the first term is maximised for $d_1^{(1)} = 0, d_3^{(1)} = d_{max}^{(1)}$ and $d_2^{(1)} = \tilde{d}^{*,(1)}$ where the latter is a solution of equation \eqref{eq-emax-ind} for the marginal prior $\tilde{\pi}_2^{(1)}$. The second term is maximised at $d_1^{(2)} = d^{*,(2)}$ which is a solution of the equation 
\begin{equation}\label{eq-help-emax}
\int_{\Theta_2^{(2)}} \frac{1}{d} - \frac{2}{d+\theta_2^{(2)}} \quad \tilde{\pi}_2^{(2)}(d \theta_2^{(2)}),
\end{equation}
in the interval $[0, d_{max}^{(2)}]$. Hence the design of the form $\xi_{\pi}^a$ maximising the Bayesian $D$-optimality criterion function is $\xi_{\pi}^{a,*} = (\xi_1^{a,*}, \xi_2^{a,*}, \mu^{a,*})$ where
\begin{equation*}
\xi_1^{a,*} = \begin{Bmatrix} 0 & \tilde{d}^{*,(1)} & d_{max}^{(1)} \\ \frac{1}{3} & \frac{1}{3} & \frac{1}{3} \end{Bmatrix}, \quad \xi_2^{a,*} = \begin{Bmatrix} d^{*,(2)} \\ 1 \end{Bmatrix}, \quad \mu^{a,*} = \begin{Bmatrix} 1 & 2 \\ \frac{3}{4} & \frac{1}{4} \end{Bmatrix}.
\end{equation*}

Following similar arguments we have that the design of the form $\xi_{\pi}^c = (\xi_1^c, \xi_2^c, \mu^c)$ maximising the Bayesian $D$-optimality criterion function is $\xi_{\pi}^{c,*} = (\xi_1^{c,*}, \xi_2^{c,*}, \mu^{c,*})$ where
\begin{equation*}
\xi_1^{c,*} = \begin{Bmatrix} d^{*,(1)} \\ 1 \end{Bmatrix}, \quad \xi_2^{c,*} = \begin{Bmatrix} 0 & \tilde{d}^{*,(2)} & d_{max}^{(2)} \\ \frac{1}{3} & \frac{1}{3} & \frac{1}{3} \end{Bmatrix}, \quad \mu^{c,*} = \begin{Bmatrix} 1 & 2 \\ \frac{1}{4} & \frac{3}{4} \end{Bmatrix},
\end{equation*}
and $d^{*,(1)}$ and $\tilde{d}^{*,(2)}$ are solutions of the equations \eqref{eq-help-emax} using the prior $\tilde{\pi}_2^{(1)}$ and \eqref{eq-emax-ind} using the prior $\tilde{\pi}_2^{(2)}$ respectively. 

For the $D$-optimality of the design $\xi_{\pi}^b = (\xi_1^b, \xi_2^b, \mu^b)$ we note that the smallest support points $d_1^{(1)}$ and $d_1^{(2)}$ of the components $\xi_1^b$ and $\xi_2^b$ must satisfy $d_1^{(1)} + d_1^{(2)} > 0$ otherwise the determinant of the information matrix vanishes. Hence there exist two possibilities corresponding to the cases $d_1^{(1)} = 0$ or $d_1^{(2)} = 0$. That is,
\begin{equation*}
\xi_1^{b_1} = \begin{Bmatrix} 0 & d_2^{(1)} \\ \frac{1}{2} & \frac{1}{2} \end{Bmatrix}, \quad \xi_2^{b_1} = \begin{Bmatrix} d_1^{(2)} & d_2^{(2)} \\ \frac{1}{2} & \frac{1}{2} \end{Bmatrix}, \quad \mu^{b_1} = \begin{Bmatrix} 1 & 2 \\ \frac{1}{2} & \frac{1}{2} \end{Bmatrix},
\end{equation*}
\begin{equation*}
\xi_1^{b_2} = \begin{Bmatrix} d_1^{(1)} & d_2^{(1)} \\ \frac{1}{2} & \frac{1}{2} \end{Bmatrix}, \quad \xi_2^{b_2} = \begin{Bmatrix} 0 & d_2^{(2)} \\ \frac{1}{2} & \frac{1}{2} \end{Bmatrix}, \quad \mu^{b_2} = \begin{Bmatrix} 1 & 2 \\ \frac{1}{2} & \frac{1}{2} \end{Bmatrix}.
\end{equation*}
Now following the same arguments as for the previous candidate designs we conclude that the designs $\xi_{\pi}^{b_1,*} = (\xi_1^{b_1,*}, \xi_2^{b_1,*}, \mu^{b_1,*})$  and $\xi_{\pi}^{b_2,*} = (\xi_1^{b_2,*}, \xi_2^{b_2,*}, \mu^{b_2,*})$ maximising the Bayesian $D$-optimality criterion function, among all designs of the corresponding form, are given by 
\begin{equation*}
\xi_1^{b_1,*} = \begin{Bmatrix} 0 & d^{*,(1)} \\ \frac{1}{2} & \frac{1}{2} \end{Bmatrix}, \quad \xi_2^{b_1,*} = \begin{Bmatrix} \tilde{d}^{*,(2)} & d_{max}^{(2)} \\ \frac{1}{2} & \frac{1}{2} \end{Bmatrix}, \quad \mu^{b_1,*} = \begin{Bmatrix} 1 & 2 \\ \frac{1}{2} & \frac{1}{2} \end{Bmatrix},
\end{equation*}
and
\begin{equation*}
\xi_1^{b_2,*} = \begin{Bmatrix} \tilde{d}^{*,(1)} & d_{max}^{(1)} \\ \frac{1}{2} & \frac{1}{2} \end{Bmatrix}, \quad \xi_2^{b_2,*} = \begin{Bmatrix} 0 & d^{*,(2)} \\ \frac{1}{2} & \frac{1}{2} \end{Bmatrix}, \quad \mu^{b_2,*} = \begin{Bmatrix} 1 & 2 \\ \frac{1}{2} & \frac{1}{2} \end{Bmatrix},
\end{equation*}
respectively, where $d^{*,(i)}$ and $\tilde{d}^{*,(i)}$ are solutions of the equations \eqref{eq-help-emax} and \eqref{eq-emax-ind} respectively using the prior $\tilde{\pi}_2^{(i)}$, $i=1,2$.

 Finally the assertion of the theorem follows by straightforward calculations comparing the Bayesian $D$-optimality criterion function of the designs $\xi_{\pi}^{a,*}$, $\xi_{\pi}^{b_1,*}$, $\xi_{\pi}^{b_2,*}$, $\xi_{\pi}^{c,*}$ in the different scenarios.

\end{proof}

\bibliographystyle{rss}
\bibliography{bayesian-commonparameters}

\end{document}